# Observation of the double quantum spin Hall phase in moiré WSe$_2$

Kaifei Kang[1*], Yichen Qiu[2], Kenji Watanabe[3], Takashi Taniguchi[3], Jie Shan[1,2,4*], and Kin Fai Mak[1,2,4*]

[1] School of Applied and Engineering Physics, Cornell University, Ithaca, NY, USA
[2] Department of Physics, Cornell University, Ithaca, NY, USA
[3] National Institute for Materials Science, Tsukuba, Japan
[4] Kavli Institute at Cornell for Nanoscale Science, Ithaca, NY, USA

[*]Email: kk726@cornell.edu, jie.shan@cornell.edu, kinfai.mak@cornell.edu

**Quantum spin Hall (QSH) insulators are a topologically protected phase of matter in two dimensions that can support non-dissipative spin transport. A hallmark of the phase is a pair of helical edge states surrounding an insulating bulk. A higher (even) number of helical edge state pairs is usually not possible in real materials because spin mixing would gap out the edge states. Multiple pairs of helical edge states have been proposed in materials with spin conservation symmetry and high spin Chern bands, but remained experimentally elusive. Here, we demonstrate a QSH phase with one and two pairs of helical edge states in twisted bilayer WSe$_2$ at moiré hole filling factor $\nu = 2$ and 4, respectively. We observe nearly quantized conductance or resistance plateaus of $\frac{h}{\nu e^2}$ at $\nu = 2$ and 4 while the bulk is insulating ($h$ and $e$ denoting the Planck's constant and electron charge, respectively). The conductance is nearly independent of out-of-plane magnetic field and decreases under an in-plane magnetic field. We also observe nonlocal transport, which is sensitive only to the in-plane magnetic field. The results agree with quantum transport of helical edge states protected by Ising spin conservation symmetry and open a promising platform for low-power spintronics.**

**Main**

Electronic materials with topological order have been intensely studied for both fundamental interest and potential device applications. Since the initial discovery of the QSH phase [1-5], researchers have asked whether a QSH insulator could support more than one pair of helical edge states [6-9], which would also increase the edge conductance (Fig. 1a). The case with two pairs of helical edge states is dubbed a 'double QSH' insulator. In principle, the double or higher QSH phase can be realized in materials with spin conservation symmetry and/or high spin Chern bands [6-15]. As such, it is robust against magnetic field along the spin quantization axis [6,7,9,12]. The phase is a close cousin of the two-dimensional topological insulators [2,3,5,16-18]. The latter are characterized by the $Z_2$ topological invariant and protected by time reversal symmetry [2,13], and hence can support only one pair of helical edge states and be destroyed by a small magnetic field [4,19-22]. While there are ample examples of two-dimensional topological or QSH insulators [HgTe and InAs/GaSb quantum wells [4,19], monolayer WTe$_2$ [20,21,23] and AB-stacked MoTe$_2$/WSe$_2$ [24,25]], experimental evidence of double or higher QSH insulators has remained elusive.

Semiconductor moiré materials [26,27], particularly twisted bilayer transition metal dichalcogenides (TMDs) [14,15,28-34], are promising to realize the double QSH phase. Exotic topological phases, including the integer and fractional Chern insulators [30-34], have been recently demonstrated in twisted bilayer MoTe$_2$ and WSe$_2$. These materials form a honeycomb moiré lattice with two sublattice sites residing in separate layers [14,15]. The topmost moiré valence bands are originated from the K and K' valley states of monolayer TMDs, which are spin-split (by about 200-500 meV) and spin-valley locked by a strong Ising spin-orbit field [35]. The field, which is nearly uniform over many moiré mini-Brillouin zones, defines a robust spin quantization axis for spin-$S_z$ conservation. Further, the continuum model with parameters from the *ab initio* calculations shows that for twist angle above 1-2 degrees [14,15], the three topmost moiré valence bands carry spin/valley-contrasting Chern numbers, $C_\uparrow = -C_\downarrow = 1$, 1 and -2, in descending order (Fig. 1b, Extended Data Fig. 1). The system is therefore a QSH insulator when the first band is filled, corresponding to moiré hole filling factor $v = 2$. The spin Chern number of the filled bands is $C_s = (C_\uparrow - C_\downarrow)/2 = 1$. The system is a double QSH insulator when the first two bands are filled at $v = 4$, and the spin Chern number of the filled bands is $C_s = 2$.

Here, we demonstrate the double QSH phase in twisted bilayer WSe$_2$ (tWSe$_2$) at $v = 4$ by combining transport, nonlocal transport and compressibility measurements. We focus on twist angle between 2.5 and 3.5 degrees to achieve strong moiré potential and weak moiré disorder at the same time. We observe incompressible phases with nearly quantized resistances of $\frac{h}{ve^2}$ and large nonlocal resistances at $v = 2$ and 4 under zero magnetic field. This is consistent with quantum transport through one and two pairs of helical edge states, respectively. These phases also exhibit negligible out-of-plane and large in-plane magnetoresistances, in agreement with helical edge states protected by spin-$S_z$ conservation symmetry.

**Device structure and uniformity**
Figure 1c illustrates the schematic device structure of dual-gated tWSe$_2$. The two gates allow independent control of the hole filling factor, $v$, and the out-of-plane electric field, $E$; the latter tunes the interlayer or sublattice potential difference [26]. The top and bottom gates are made of palladium (Pd) and graphite electrodes, respectively. The dielectric in both gates is hexagonal boron nitride (hBN). To achieve good electrical contacts to the tWSe$_2$ channel, we use platinum (Pt) contact electrodes and additional Pd contact gates to heavily dope tWSe$_2$ in the contact region.

Figure 1d shows an optical image of a 3.0-degree tWSe$_2$ transport device (D1). The twist angle is calibrated from the Hall density measurement (Extended Data Fig. 2). Because twist angle disorder is generally difficult to eliminate in small-angle-twisted homobilayers [26], we identify a small area (shaded red) with relatively uniform twist angle and focus on this area in the measurement (see Extended Data Fig. 3 for characterization of device uniformity). Seven contacts (labeled 1-7) exhibit low contact resistances (about 10 kΩ at 1.5 K); four of them (2, 3, 6, 7) make contact to the uniform region of the channel and are used as voltage probes. Unless otherwise specified, results are presented for measurements on D1 at temperature $T = 1.5$ K and under zero magnetic field. Similar results are observed in second transport device (D2, 3.5-degree tWSe$_2$) which are included in Extended Data Fig. 4. Details on the device fabrication, electrical measurements and band structure calculations are provided in Methods.

**Local transport**
We study local transport in device D1 by biasing a current between electrode 1 and 4 and measuring the voltage drop between electrode 2 and 3 and between 7 and 6 (inset of Fig. 2c). Electrode (2, 3) and (7, 6) are on two opposing sides of the channel and have similar separations of about 1.5 μm. The nearly identical resistances, $R_{14,23}$ and $R_{14,76}$, under different conditions demonstrate the device uniformity (Fig. 2c, 2d and Extended Data Fig. 5). The bias current is nearly equally partitioned on the two sides of the channel, which also leads to negligible longitudinal-transverse mixing (Extended Data Fig. 3b).

Figure 2a shows resistance $R_{14,23}$, or simply $R$, as a function of electric field and filling factor. Data is not available for certain parameter regions due to gate limitations. There are two high resistance islands, around $\nu = 2$ and 4, which are symmetric about $E = 0$ (the value of the field can deviate slightly from zero due to small structure/device asymmetries, and can vary after a thermal cycle). Figure 2c is a linecut along $E \approx 0$. The resistance plateau/peak around $\nu = 2$ and 4 is close to the quantized value of $\frac{h}{\nu e^2}$ (about 10-15% higher). Figure 2d illustrates linecuts along $\nu = 2$ and 4. The resistance also shows plateaus/peaks in electric field around $E = 0$ with nearly quantized values.

Figure 2b illustrates the penetration capacitance $C_p/C_g$, normalized by the gate-to-gate geometric capacitance $C_g$, of a 3.0-degree tWSe$_2$ capacitance device (D3). The penetration capacitance measures the bulk electronic compressibility; the incompressible states correspond to peaks in $C_p/C_g$ [36]. Both the $\nu = 2$ and 4 states are incompressible around $E = 0$. We determine their charge gap to be about 4 and 1.5 meV, respectively, from the gate-voltage dependence of $C_p/C_g$ (Methods). Given that the bulk is insulating and the thermal energy is much smaller than the bulk charge gaps (see Fig. 5), the nearly quantized resistance/conductance observed in local transport suggests the existence of edge channels.

**Nonlocal transport**
We perform the nonlocal transport measurement to verify the existence of edge channels [37]. In contrast to bulk transport, edge channel transport necessarily leads to nonlocal resistance, which is also expected to be quantized if the coherence length of the edge states is long compared to the edge channel length. As shown in the inset of Fig. 3a, we bias a current between electrode 3 and 6, and measure the voltage drop between 2 and 1. The edge channel length is about 5 μm.

Figure 3a shows resistance $R_{36,21}$, or nonlocal resistance $R_{NL}$, in device D1 as a function of filling factor and electric field. Large $R_{NL}$ is observed near $\nu = 2$ and 4 in the small electric-field regime. Specifically, the maximum nonlocal resistance is about 1 kΩ and 400 Ω at $\nu = 2$ and 4, respectively. In contrast, the background $R_{NL}$ is $\lesssim 10$ Ω even in the region of $\nu < 1$ and $E \approx 0$, where the local resistance $R$ reaches about 100 kΩ (Fig. 2a);

the result shows that the nonlocal resistance peaks at $\nu = 2$ and 4 are not experimental artifacts from a highly insulating bulk channel. The negligible nonlocal resistance in the background is consistent with bulk transport and exponential suppression of bulk transport in the nonlocal measurement geometry (Methods). The large nonlocal resistances at $\nu = 2$ and 4 relative to their respective local resistances (up to about 10 %) support the importance of edge channel transport at these filling factors.

**Magnetic-field responses**
We examine the magnetic-field effect on the edge channels at $\nu = 2$ and 4. We focus on the special case of $E \approx 0$ for which the nonlocal response is the strongest (see Extended Data Fig. 6 for the complete dependence on $\nu$ and $E$). Spin Chern bands are also expected only in the small electric-field regime where the moiré bands are layer hybridized [14,15,30-33]. Figure 4a, b illustrate the filling-factor dependence of resistance $R$ under representative out-of-plane ($B_\perp$) and in-plane magnetic ($B_\parallel$) fields, respectively. (The sample was warmed up and reloaded into the cryostat to alter the magnetic-field orientation; this process introduced small changes to the sample and the resistances under zero magnetic field in Fig. 4a, b.) The resistance plateaus around $\nu = 2$ and 4 are nearly independent of $B_\perp$, but increase with $B_\parallel$. Meanwhile, the resistance in the compressible regions away from $\nu = 2$ and 4 shows opposite behavior. We describe the field effect at $\nu = 2$ and 4 in terms of conductance $G/G(0\,T)$, normalized by its zero field value $G(0\,T)$. Figure 4c, d show the in-plane field dependence of $G/G(0\,T)$ at representative temperatures for $\nu = 2$ and 4, respectively. All curves are peaked at zero field and decay with field amplitude. At 1.5 K, the normalized conductance saturates to ~ 40% at $\nu = 2$ and ~ 60% at $\nu = 4$. As temperature increases, the normalized conductance peak expands in field and saturates at higher values.

Under small in-plane fields, local transport shows a thermal activation behavior (Extended Data Fig. 7). We estimate the activation energy (transport gap) from the temperature dependence of resistance or conductance for each field. The transport gap increases linearly with $B_\parallel$ for small fields and saturates above 0.2 T (Fig. 4e). The saturated transport gap is about 0.13 meV ($\nu = 2$) and 0.05 meV ($\nu = 4$). In comparison, the bulk charge gaps determined from the compressibility measurement do not depend on $B_\parallel$ up to 14 T--the highest field available in this study (Fig. 4f).

Figure 3b, c illustrate the magnetic-field dependence of $R_{NL}$. Similar to local transport, nonlocal transport is weakly dependent on $B_\perp$ and is sensitive to $B_\parallel$ at $\nu = 2$ and 4 (Fig. 3c). The in-plane magnetic field suppresses nonlocal transport; $R_{NL}$ at 2 T is reduced to about 25% of its value at zero field for both $\nu = 2$ and 4 (Fig. 3b, c).

**Discussions**
The results above are inconsistent with band insulating states at $\nu = 2$ and 4, which would not support edge channel transport and large in-plane magnetoresistance (due to spin-$S_z$ conservation). Instead, the results are consistent with a QSH insulator with one and two pairs of helical edge states protected by spin-$S_z$ conservation symmetry at $\nu = 2$ and 4, respectively, in tWSe$_2$. The penetration capacitance shows that these states are insulators (Fig. 2b) and the bulk charge gaps are independent of $B_\parallel$ (Fig. 4f). The latter is a characteristic of materials with Ising-like spins. Yet the bulk charge gaps in Fig. 4f are distinct from the $B_\parallel$-dependent transport gaps in Fig. 4e. The result can only be explained by the presence of edge channel transport at $\nu = 2$ and 4.

The large nonlocal resistance at $\nu = 2$ and 4 further supports the existence of edge channel transport. The nearly quantized resistance $R \approx \frac{h}{\nu e^2}$ with plateaus/peaks in both filling factor and electric field (Fig. 2c, d) indicates that the edge states are helical and there are one and two pairs of helical edge states at $\nu = 2$ and 4, respectively. This is consistent with the continuum model band structure calculations (Extended Data Fig. 1), which predict spin Chern number $C_s = 1$ and 2 for $\nu = 2$ and 4, respectively [15]. The less pronounced resistance plateau at $\nu = 4$ compared to that at $\nu = 2$ arises from stronger thermally activated bulk transport at $\nu = 4$, which is consistent with its smaller bulk charge gap. The 10-15% higher resistance than the quantized value for both states suggests that the coherence length of the helical edge states is comparable to or shorter than 1.5 μm (the separation between the probe electrodes). The smaller nonlocal resistance compared to the expected quantized value from Landauer-Buttiker analysis ($\frac{2h}{5\nu e^2} = 5.16$ and 2.58 kΩ for $\nu = 2$ and 4, respectively), especially for $\nu = 4$, is again a manifestation of finite bulk conduction between electrodes 3 and 6 in Fig. 3a inset. The nonlocal resistance is much more susceptible to bulk conduction than local resistance because the bulk channel (localized in between electrodes 3 and 6, separated by about 2 μm) is much shorter than the edge channel (nearly 10 μm through electrodes 3-2-1-7-6) in the nonlocal measurement geometry. The sensitivity of nonlocal measurements to bulk conduction is further demonstrated by the suppressed (rather than enhanced) $R_{NL}$ under $B_\parallel$ (Fig. 3b), which gaps out the edge states (see Fig. 4e and below). In contrast, the enhanced local resistance $R$ under $B_\parallel$ (Fig. 4b) shows the dominance of edge channel transport in local measurements (Fig. 2c inset).

The results above also show that the energy scale of the QSH phase is determined by the bulk charge gap (about 4 meV and 1.5 meV at $\nu = 2$ and 4, respectively). This is further supported by the temperature dependence of $R$ and $R_{NL}$ in Fig. 5. The resistance plateaus at $\nu = 2$ and 4 are moderately smeared at 20 K, whereas $R$ rapidly increases with increasing temperature for the compressible regions away from $\nu = 2$ and 4 (Fig. 5a). On the other hand, $R_{NL}$ at $\nu = 2$ and 4 decreases monotonically with increasing temperature in the same temperature window (Fig. 5b). This again reflects the stronger influence of bulk conduction in nonlocal transport.

The negligible out-of-plane and strong in-plane magnetic-field effects on transport (Fig. 3, 4) show that the helical edge states are protected by spin-$S_z$ conservation symmetry. Specially, $B_\perp$ is along the spin quantization axis and does not affect the helical edge states (although it can affect the bulk transport in the compressible regions through orbital effects). Contrarily, whereas $B_\parallel$ has no effect on the bulk transport, it can induce spin mixing and backscattering, open a gap, and localize the edge states, as shown in Ref. [22]. This picture is consistent with the emergence of transport gap under an in-plane field (Fig. 4e). It is also consistent with the suppression of nonlocal transport under $B_\parallel$ (Fig. 3b, c). However, we note that edge channel transport remains relevant even under high in-plane fields (Extended Data Fig. 8). In particular, the nonlocal resistance at $v = 2$ and 4 under $B_\parallel = 2$ T (100 - 300 Ω) is still substantially higher than the background (≤ 10 Ω) (Fig. 3b).

**Concluding remarks**

By combining penetration capacitance and transport measurements, we have demonstrated that a QSH insulator can have more than one pair of helical edge states. Specifically, tWSe$_2$ is a double QSH insulator at hole moiré filling factor $v = 4$. The new phase is endowed by the presence of spin conservation symmetry and high spin Chern bands in the material [14,15]. It is therefore expected in other twisted bilayer TMDs (such as MoTe$_2$) with appropriate twist angles that share the same electronic properties. The continuum model band structure calculations [15] show that for very small twist angles (~ 1 degree or less), only the two topmost moiré valence bands carry spin/valley-contrasting Chern numbers, $C_\uparrow = -C_\downarrow = 1$ and -1 (Extended Data Fig. 1). In this case, a trivial band insulator, not a double QSH insulator, is expected at $v = 4$. The double QSH insulators, which support two pairs of helical edge states and double edge conduction, open a promising platform for low-power spintronics beyond two-dimensional topological insulators [6-9]; they also provide an opportunity to realize a bosonic QSH insulator in future studies [38].

**Methods**

**Device fabrication and twist angle calibration**

The tWSe$_2$ dual-gate devices were fabricated using the tear-and-stack and layer-by-layer transfer method [39,40]. In short, flakes of monolayer WSe$_2$, hexagonal boron-nitride (hBN) and few-layer graphite were first exfoliated and identified by their optical contrast. They were transferred using a thin film of PC (polycarbonate) on PDMS (polydimethylsiloxane). We sequentially picked up a hBN flake, part of monolayer WSe$_2$, the rest of the monolayer at a small twisting angle, another hBN flake and a graphite flake. The finished hBN/WSe$_2$/WSe$_2$/hBN/Graphite stack was released at a temperature of 180°C onto a Si/SiO2 substrate with pre-patterned platinum electrodes in the Hall bar geometry. We then defined a rectangular device channel by standard e-beam lithography, metallization and reactive ion etching. For the capacitance device D3, which has nearly symmetric top and bottom gates, the finished stacks were released onto pre-patterned quartz substrates to reduce the parasitic capacitance background [36].

We calibrate the hole doping density in tWSe$_2$ by the small-field (0.4 T) Hall effect measurement (Extended Data Fig. 2). The measured Hall density allows us to determine the moiré density and the twist angle of the device based on the positions of the $v = 2$ and 4 insulating states. For the device in the main text, the twist angle is 3.0-degree, which is very close to the target twist angle 2.9-degree. We can also calibrate the hole doping density by using the parallel-plate capacitor model, $n = \frac{\varepsilon\varepsilon_0 V_{tg}}{d_{tg}} + \frac{\varepsilon\varepsilon_0 V_{bg}}{d_{bg}}$. Here $V_{tg}$ and $V_{bg}$ denote, respectively, the top and bottom gate voltages, $\varepsilon_0$ is the vacuum permittivity, $\varepsilon \approx 3$ is the hBN dielectric constant, and $d_{tg}$ and $d_{bg}$ are, respectively, the top and bottom hBN gate dielectric thicknesses. For the device in the main text, $d_{tg}$ and

$d_{tg}$ were measured by the atomic force microscopy to be 5.8 nm and 8.4 nm, respectively. We then calibrate the filling factor $\nu = 2$ and 4 for the insulating states, and obtain the twist angle to be 2.7-degree, which is in good agreement with the value obtained from the Hall density measurement. We further define the out-of-plane electric field as $E = \frac{1}{2}(\frac{V_{tg}}{d_{tg}} - \frac{V_{bg}}{d_{bg}})$.

**Electrical transport measurements**

Electrical transport measurements were performed in a closed-cycle $^4$He cryostat (Oxford TeslatronPT) equipped with a superconducting magnet up to 14 T. Low-frequency (< 23 Hz) lock-in techniques were used to measure the sample resistance under a small bias current of less than 10 nA to avoid sample heating and/or high bias effects. We biased the device symmetrically along the major axis of the Hall bar to ensure a uniform current flow. Both the source-drain current and the voltage drop at the probe electrode pairs were recorded. Voltage pre-amplifiers with large input impedance (100 MΩ) were used to measure the sample resistance up to about 10 MΩ.

**Nonlocal resistance measurements**

The device in the main text is approximately a symmetric Hall bar device. The nonlocal measurement geometry is shown in the inset of Fig. 3a. With a bias current along electrodes 3 and 6 and voltage drop measured between 2 and 1, the non-local resistance due to bulk transport is suppressed by a factor more than $\rho \exp(-\pi l/w)$ [Ref.[41]]. Here $\rho$, $l$ and $w$ denote the 2D sheet resistivity of tWSe$_2$, the center-to-center distance between the source-drain pair (3-6) and the probe pair (2-1) and the width of the Hall bar, respectively. We have $w \approx 1.5$ μm and $l \approx 4$ μm in our device; the non-local resistance is approximately 0.02% of the bulk resistivity $\rho$ if there were no edge state transport. This estimate is consistent with the negligible non-local resistance at fillings away from $\nu = 2$ and 4.

For an ideal QSH insulator, in which the bulk has infinite resistivity and the edge states are perfectly helical without backscattering, the nonlocal resistance for the measurement geometry in Fig. 3a is simply given by the resistance quantum $\frac{h}{e^2}$ multiplied by the ratio of the current through the path 3-2-1-7-6 to the total source-drain current. The latter can be calculated by counting the number of edges connecting each adjacent electrode pairs with each contributing a resistance quantum. We get $R_{NL} = \frac{2h}{5\nu e^2} = 5.16$ kΩ and 2.58 kΩ for $\nu = 2$ and 4, respectively.

The experimental $R_{NL}$ is substantially smaller than these ideal values (especially for $\nu = 4$) because of the combined effects of remnant bulk conduction and the finite coherence length of the helical edge states. At high $B_\parallel$ or $T$, the edge state transport is strongly suppressed and the bulk transport dominates. We expect the current density to concentrate between the source and drain electrodes (3-6) in this bulk transport regime. We also expect a smooth evolution of the current flow pattern between the edge-dominate and the bulk-dominate limits as $B_\parallel$ or $T$ increases.

## Capacitance measurements

The capacitance measurements were performed in the Oxford TeslatronPT cryostat. Details have been reported in a recent study on MoSe$_2$/WS$_2$ moiré heterobilayers [36]. In short, a commercial high electron mobility transistor (HEMT, model FHX35X) was employed as the first-stage amplifier [42]. The HEMT was connected to the sample on the same amorphous quartz chip. The differential penetration capacitance, $C_p$, was measured by applying an AC voltage (10 mV in amplitude and 3–5 kHz in frequency) to the top gate and collecting the charge current from the bottom gate through the HEMT by lock-in techniques.

We determined the charge gaps of the $\nu = 2$ and 4 states from the penetration capacitance. The penetration capacitance can be expressed as $C_p \approx \frac{C_t \cdot C_b}{C_t + C_b + C_Q}$, where $C_t$ and $C_b \approx C_t$ are the geometric top-gate and back-gate capacitances, respectively, and $C_Q$ is the quantum capacitance. The thermodynamic gap is reflected as a chemical potential jump, $\Delta\mu$, which is obtained as $\int \frac{C_p}{C_t} dV_{bg} \approx \int \frac{C_p}{2C_g} dV_{bg}$ ($C_g$ is the gate-to-gate geometric capacitance). Here the integration spans the range of the capacitance peaks corresponding to the $\nu = 2$ and 4 insulating states.

## Continuum model

The moiré band structure of small-angle-twisted WSe$_2$ bilayers was calculated using a continuum model following Ref. [14]. The Hamiltonian for the K valley states of WSe$_2$ is given by

$$H_K = \begin{pmatrix} \frac{\hbar^2 k^2}{2m^*} + \Delta_t(\mathbf{r}) + \frac{V_z}{2} & \Delta_T(\mathbf{r}) \\ \Delta_T^\dagger(\mathbf{r}) & \frac{\hbar^2 k^2}{2m^*} + \Delta_b(\mathbf{r}) - \frac{V_z}{2} \end{pmatrix} \quad (1)$$

Here $\frac{\hbar^2 k^2}{2m^*}$ is the kinetic energy with $k$ and $m^*$ denoting the crystal momentum and effective mass; $\Delta_{t,b}(\mathbf{r}) = 2V \sum_{j=1,3,5} \cos(\mathbf{g}_j \cdot \mathbf{r} \pm \psi)$ are, respectively, the spatially ($\mathbf{r}$) dependent moiré potential for the top ($t$) and bottom ($b$) layers with depth $V$ and phase $\psi$ expanded in harmonic components ($\mathbf{g}_j$ is the moiré reciprocal wavevector); $\Delta_T(\mathbf{r}) = w\left(1 + e^{i\mathbf{g}_2 \cdot \mathbf{r}} + e^{i\mathbf{g}_3 \cdot \mathbf{r}}\right)$ is the interlayer tunneling with amplitude $w$; and $V_z$ is the interlayer or sublattice potential difference induced by an out-of-plane electric field. In the calculation, we used $m^* = 0.43\, m_0$ ($m_0$ denoting the free electron mass) and $(V, \psi, w) = (9 meV, -128°, -18 meV)$ [14]. The Hamiltonian was diagonalized using the plane-wave method and cut off at the 5$^{th}$ shell of the moiré Brillouin zone. The spin/valley Chern number of each moiré band was obtained by integrating the Berry curvatures within the first Brillouin zone. The Hamiltonian for the -K valley states is a time-reversed copy of Eqn. (1).

## Figures

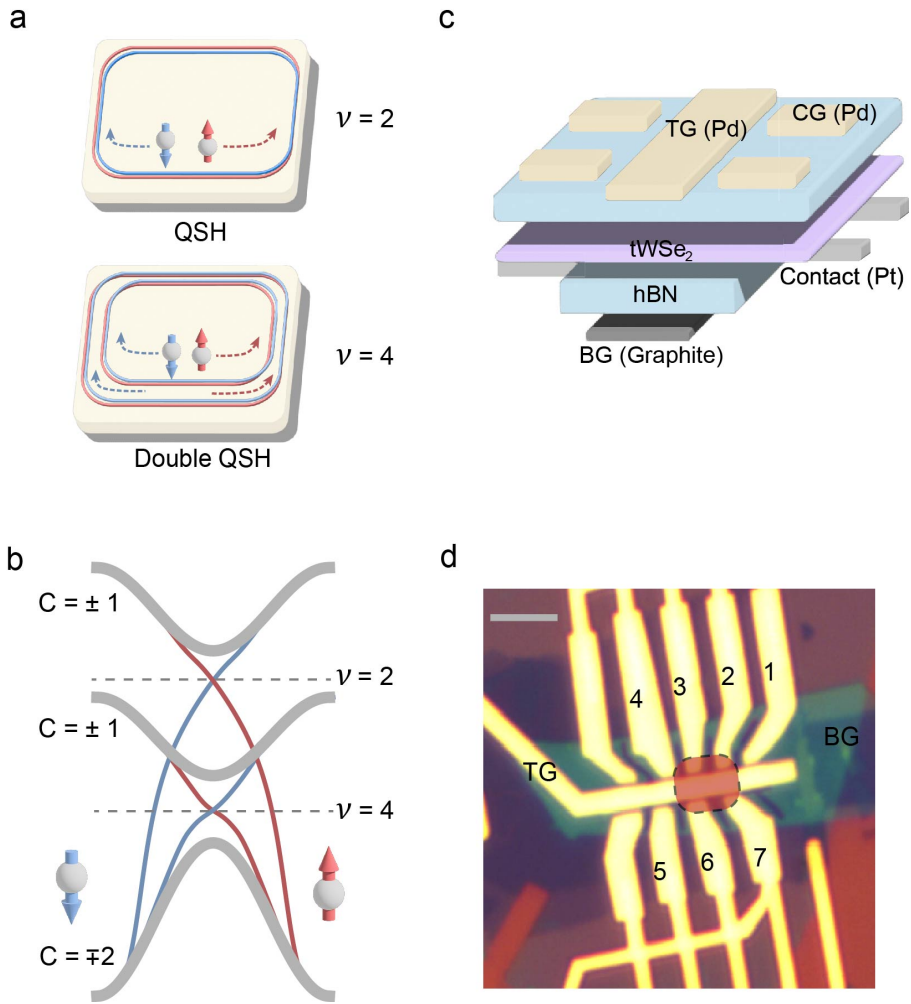

**Figure 1 | Double quantum spin Hall (QSH) phase in tWSe$_2$. a,** Schematic of a QSH insulator with one (top) and two (bottom) pairs of helical edge states. **b,** Schematic band structure of tWSe$_2$ for the K (K') valley states. The three topmost moiré valence bands carry spin/valley-contrasting Chern numbers, $C_\uparrow = -C_\downarrow = 1$, 1 and -2. The dashed lines denote the Fermi level at hole filling factor ν = 2 and 4. The red and blue lines in **a**, **b** represent the counter propagating edge modes with up- and down-spins. **c,** Schematic device structure. Purple: tWSe$_2$ channel; light blue: hBN dielectric; grey: platinum (Pt) contact electrodes; black: few-layer graphite bottom gate; yellow: palladium (Pd) top gate and contact gates. The contact gates heavily hole-dope the tWSe$_2$ regions near the Pt electrodes to achieve low contact resistances. **d,** Optical image of device D1. Electrodes (1-7) have low contact resistances. The red shaded area denotes the uniform moiré region. The scale bar is 5 μm.

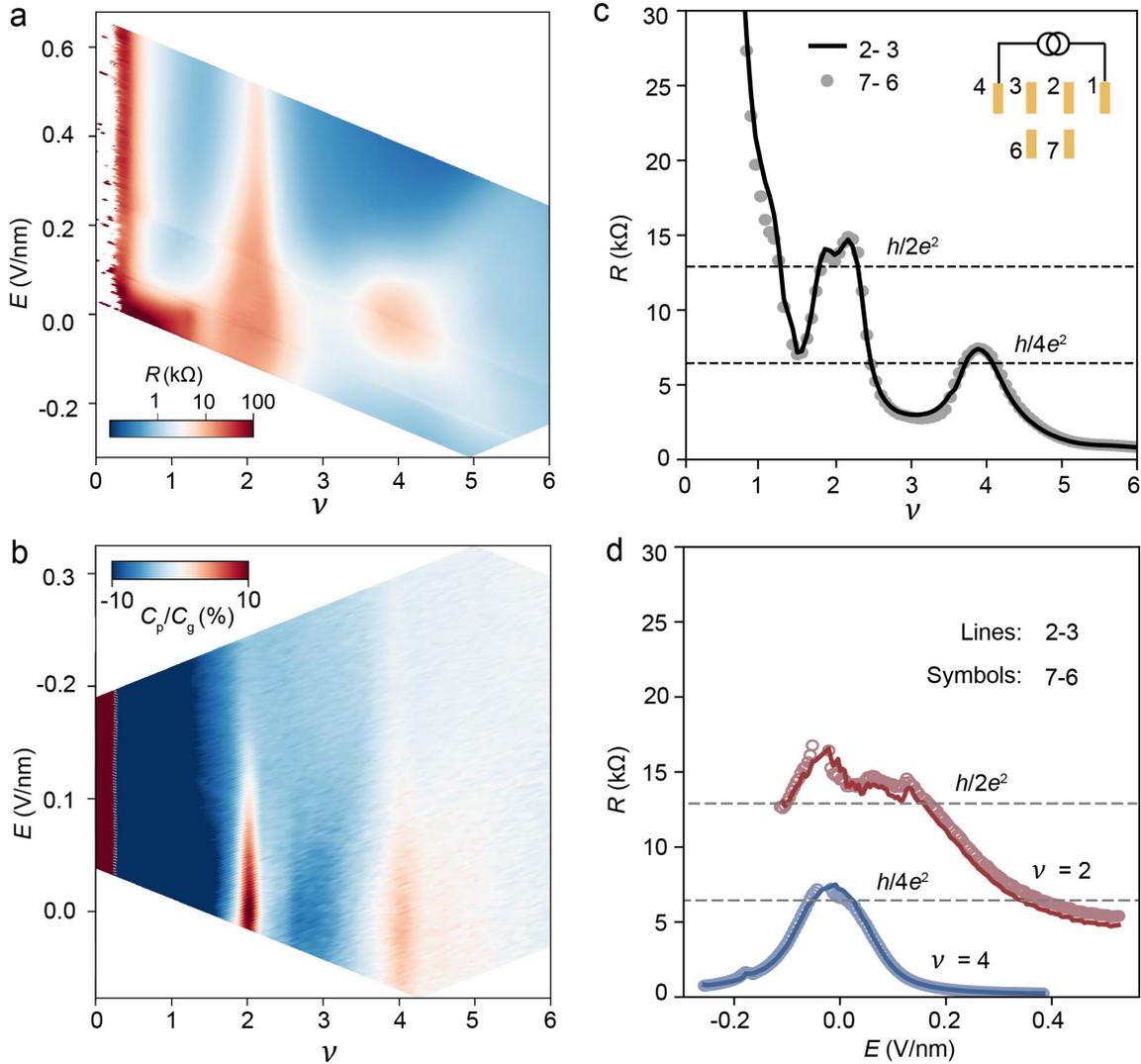

**Figure 2 | Local transport. a, b,** Longitudinal resistance $R$ or $R_{14,23}$ (**a**) and normalized penetration capacitance $C_p/C_g$ (**b**) as a function of out-of-plane electric field ($E$) and hole filling factor ($\nu$). **c,** Filling-factor dependence of $R$ at $E = 0.03$ V/nm. Inset: current is bias between electrode 1 and 4; voltage drop is probed between electrode pair (2, 3) or (7, 6). **d,** Electric-field dependence of $R$ at $\nu = 2$ (red) and 4 (blue). The dashed lines in **c, d** denote the quantized resistances $h/2e^2$ and $h/4e^2$. All measurements are at $T = 1.5$ K under zero magnetic field.

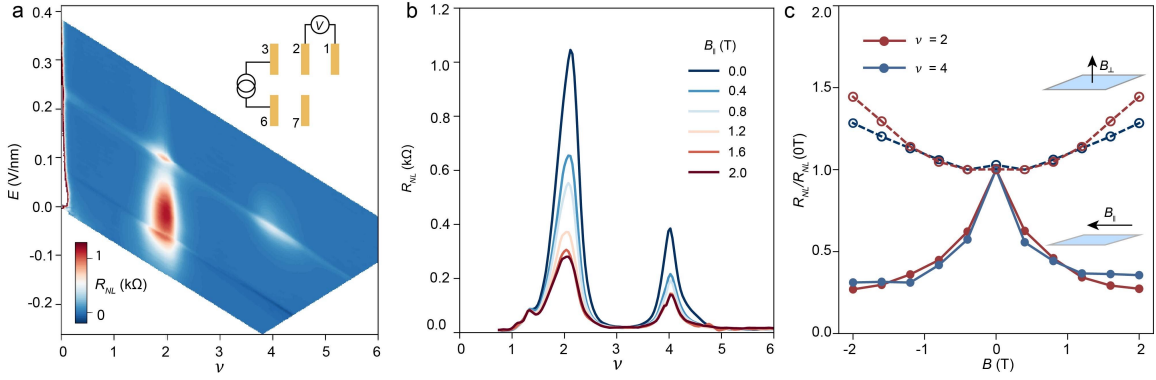

**Figure 3 | Nonlocal transport. a,** Nonlocal resistance $R_{NL}$ as a function of out-of-plane electric field ($E$) and hole filling factor ($\nu$). Inset: current is bias between electrode 3 and 6; voltage drop is probed between electrode 1 and 2. **b,** Filling-factor dependence of $R_{NL}$ at $E = 0$ V/nm under varying in-plane magnetic fields between 0 and 2 T. **c,** Out-of-plane (dashed lines and empty dots) and in-plane (solid lines and dots) magnetic-field dependences of normalized nonlocal resistance, $R_{NL}/R_{NL}(0\,T)$, at $\nu = 2$ (red) and 4 (blue). All measurements are at $T = 1.5$ K.

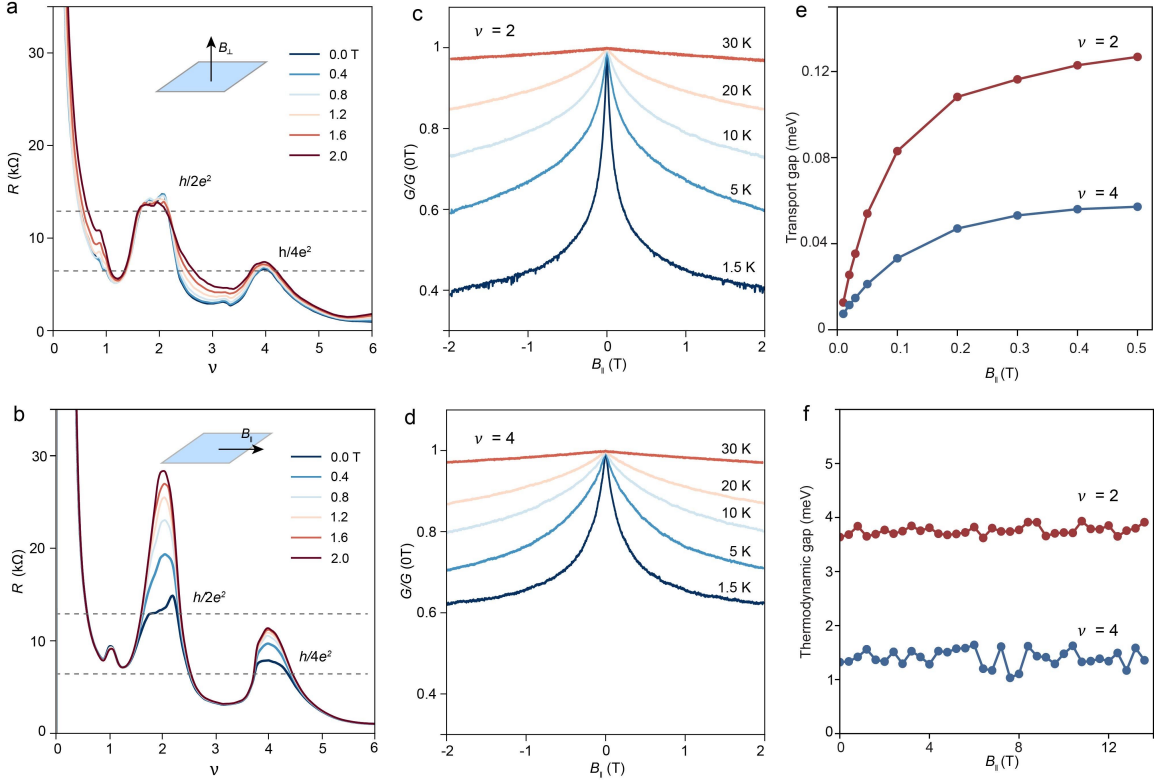

**Figure 4 | Magnetic-field responses. a, b,** Filling-factor dependence of local resistance at $T = 1.5$ K under varying out-of-plane (**a**, $E = 0.03$ V/nm) and in-plane (**b**, $E = 0.01$ V/nm) magnetic fields between 0 and 2 T. The dashed lines denote the expected quantized value of $h/\nu e^2$ at $\nu = 2$ and 4. **c, d,** In-plane magnetic-field dependence of normalized conductance, $G/G(0\text{T})$, at $\nu = 2$ (**c**) and $\nu = 4$ (**d**) at selected temperatures from 1.5 to 30 K. **e,** Transport gap or activation energy as a function of in-plane magnetic field, extracted from the temperature dependence of resistance at each field for small fields (Extended Data Fig. 7). **f,** Thermodynamic or bulk charge gap as a function of in-plane magnetic field up to 14 T, extracted from the penetration capacitance measurement. In **e**, **f**, red and blue denote $\nu = 2$ and 4, respectively. Symbols are values extracted from experiment and lines are guides to the eye.

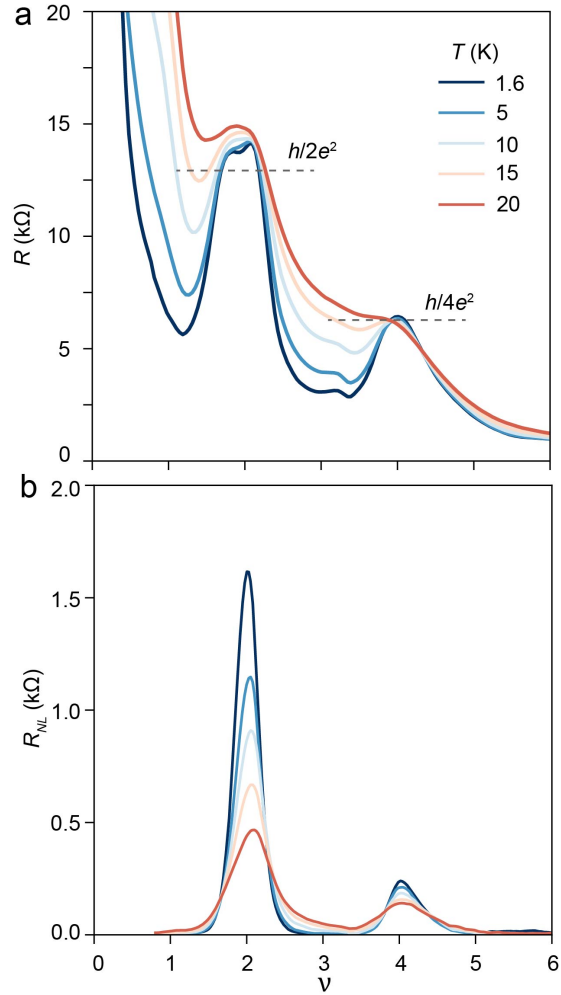

**Figure 5 | Temperature dependence. a, b,** Filling-factor dependence of local ($R$, **a**) and nonlocal ($R_{NL}$, **b**) resistances at $E = 0$ V/nm and varying temperatures from 1.6 to 20 K. The dashed lines in **a** denote the expected quantized value of $h/\nu e^2$ at $\nu = 2$ and 4.

**Extended Data Figures**

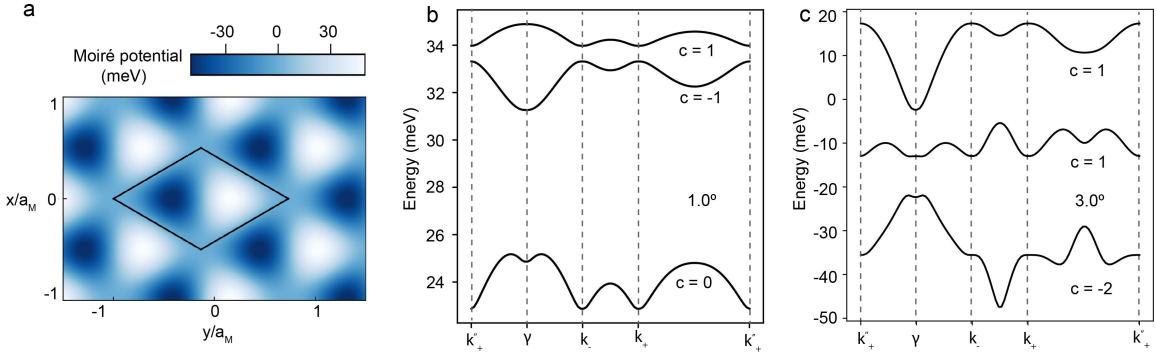

**Extended Data Figure 1 | Electronic band structure of tWSe$_2$. a,** Spatial dependence of moiré potential, where $x$ and $y$ denote the in-plane spatial coordinates in units of moiré period $a_M$. The potential maximum and minimum are associated with two different layers. Together they form a honeycomb lattice. The black lines mark a moiré unit cell. **b, c,** Continuum model moiré band structure of the K valley states under zero interlayer potential difference for 1.0-degree (**b**) and 3.0-degree (**c**) tWSe$_2$. The three topmost moiré valence bands are illustrated. Chern number $c$ is nonzero for the first two (**b**) and three (**c**) bands. The moiré bands from the K' valley states are time-reversal copies and carry opposite Chern numbers.

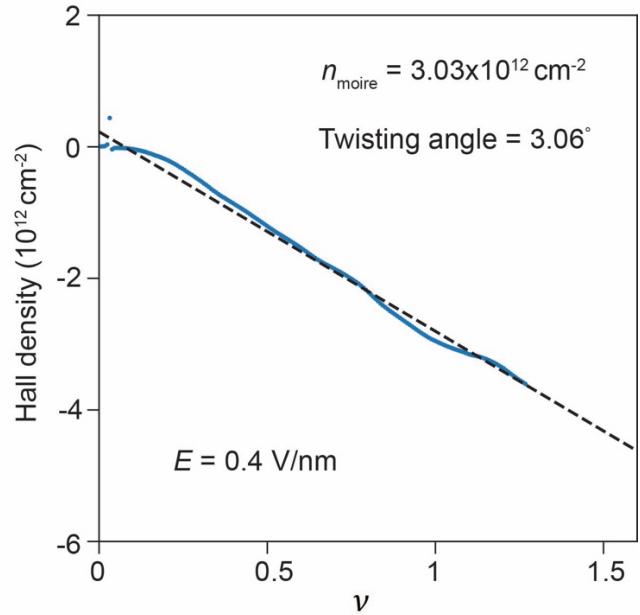

**Extended Data Figure 2 | Twist angle calibration by the Hall density measurement (D1).** Filling-factor dependence of small-field ($B_\perp = 0.4$ T) Hall density. A high electric field ($E = 0.4$ V/nm) is applied so that the $\nu = 1$ state is metallic. A linear fit (dashed line) to the data (solid line) yields moiré density $(3.03 \pm 0.03) \times 10^{12}$ cm$^{-2}$ and twist angle $(3.06 \pm 0.03)$ degrees.

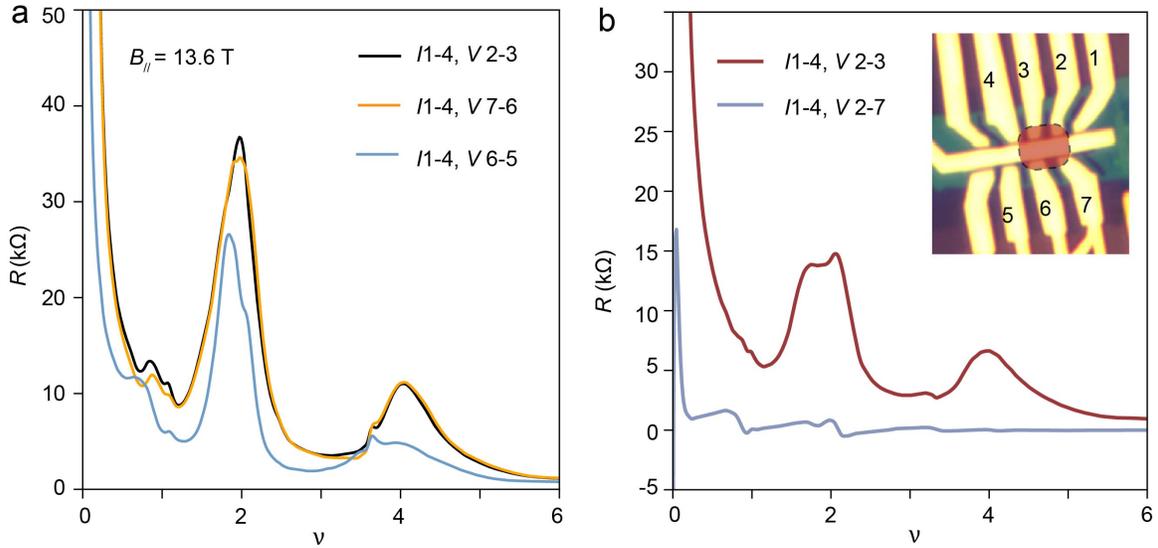

**Extended Data Figure 3 | Characterization of device uniformity (D1). a,** Filling-factor dependence of resistance. Current is biased between electrode 1 and 4; voltage drop is probed between electrode pair (2, 3), (7, 6) and (6, 5). A large in-plane magnetic field ($B_\parallel = 13.6$ T) is applied to localize the edge states in order to better access the bulk transport. Probe pair (2, 3) and (7, 6) yield nearly identical resistances. Probe pair (6, 5) measures resistance peaks down shifted in filling factor, indicating a smaller moiré density for the region. **b,** Filling-factor dependence of longitudinal (red) and transverse (blue) resistances under zero magnetic field. The transverse resistance is nearly an order of magnitude smaller than the longitudinal resistance, demonstrating weak longitudinal-transverse mixing in the uniform moiré region. Inset: optical image of the device with electrode (1-7) labeled. Area with uniform moiré is shaded red. All measurements are at 1.5 K.

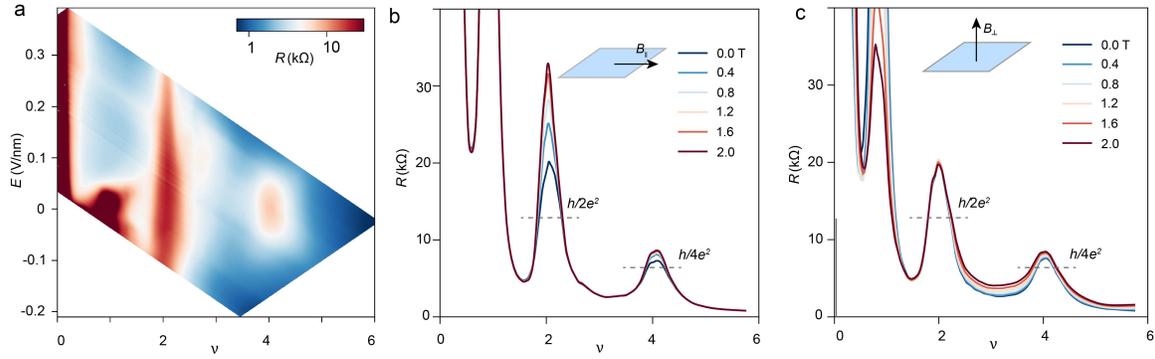

**Extended Data Figure 4 | Results from device D2 (3.5-degree tWSe$_2$). a,** Local resistance ($R$) as a function of out-of-plane electric field ($E$) and hole filling factor ($\nu$). **b, c,** Filling-factor dependence of $R$ under varying in-plane (**b**, $E = 0.02$ V/nm) and out-of-plane (**c**, $E = 0.02$ V/nm) magnetic fields between 0 and 2 T. The dashed lines in **b**, **c** denote the expected quantized value of $h/\nu e^2$ at $\nu = 2$ and 4. All measurements are at 1.5 K.

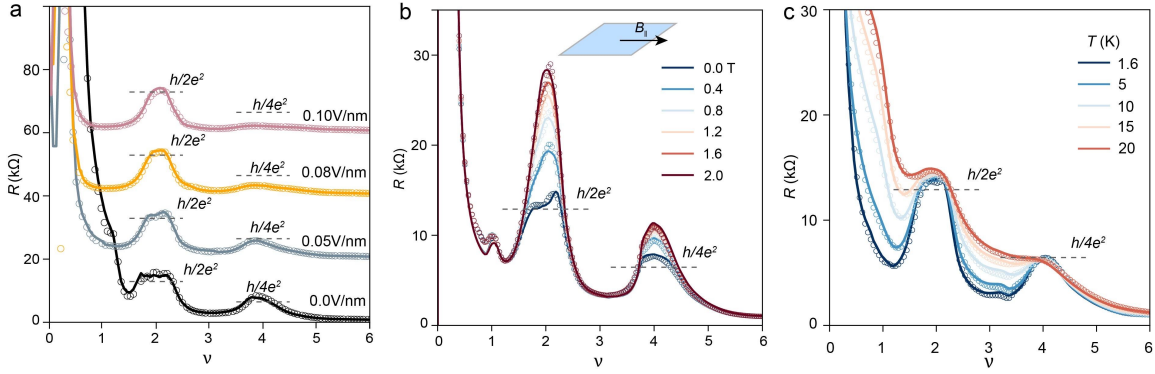

**Extended Data Figure 5 | Extra local transport results (D1). a-c,** Filling-factor dependence of resistance $R_{14,23}$ (lines) and $R_{14,76}$ (symbols) at varying electric fields (**a**), in-plane magnetic fields (**b**) and temperatures (**c**). The temperature is fixed at 1.5 K in **a**, **b**; the magnetic field is zero in **a**, **c**; and the out-of-plane electric field is fixed at $E = 0.01$ V/nm in **b**, **c**. The curves in **a** are vertically shifted for clarity. The dashed lines denote the expected quantized value of $h/\nu e^2$ at $\nu = 2$ and 4.

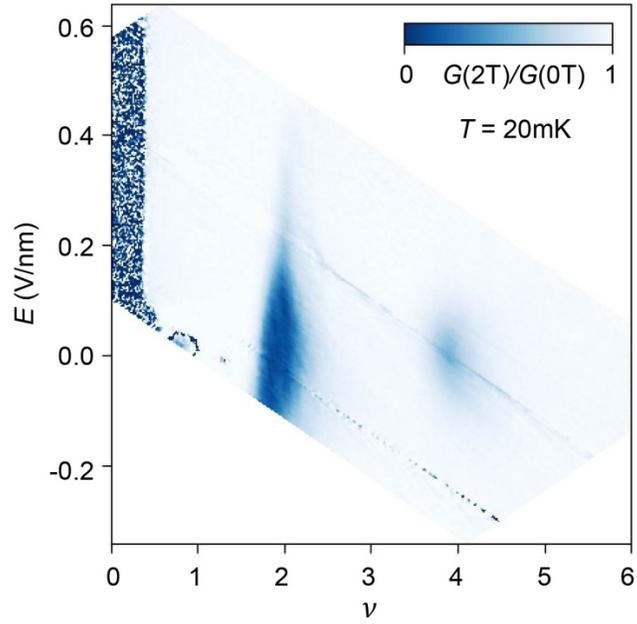

**Extended Data Figure 6 | In-plane magneto-conductance (D1).** Normalized conductance at $B_\parallel = 2$ T by the zero-field conductance, $G(2T)/G(0T)$, as a function of out-of-plane electric field ($E$) and hole filling factor ($\nu$) at lattice temperature of 20 mK. At zero electric field, the conductance is suppressed by the in-plane magnetic field to about 10% near $\nu = 2$ and 50% at $\nu = 4$.

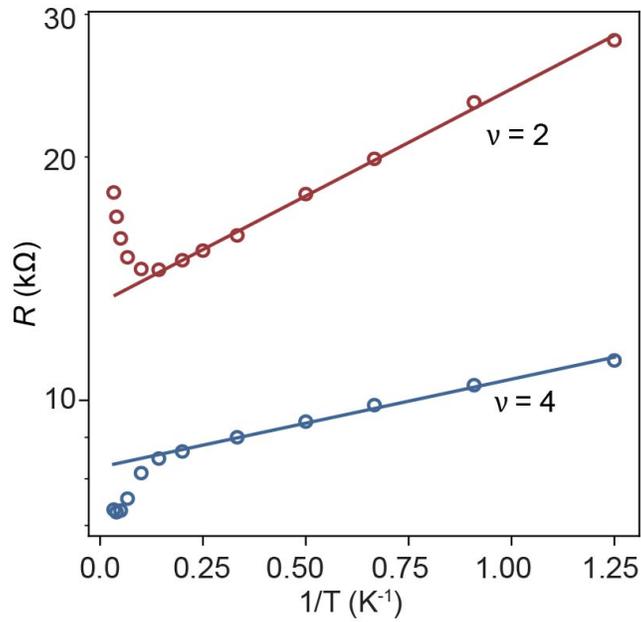

**Extended Data Figure 7 | Thermally activated edge transport under an in-plane magnetic field (D1).** Arrhenius plot of resistance $R$ at $\nu = 2$ and 4 under $B_\parallel = 0.2$ T and $E = 0$ V/nm. Symbols: experiment; solid lines: thermal activation fits with activation energy of 108 and 47 µeV for $\nu = 2$ and 4, respectively.

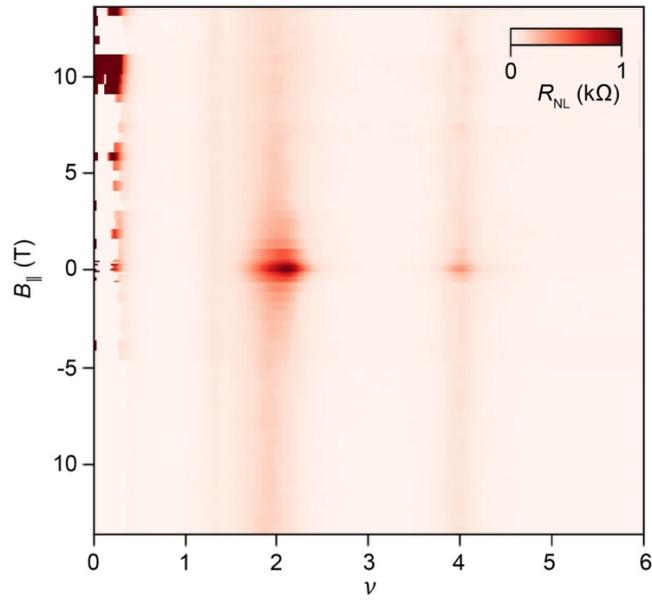

**Extended Data Figure 8 | Nonlocal transport under high magnetic fields (D1).** Nonlocal resistance as a function of filling factor and in-plane magnetic field up to 13.6 T at $E \approx 0$ V/nm and $T = 1.5$ K. The nonlocal resistance at $\nu = 2$ and 4 decreases rapidly with increasing in-plane magnetic field, but remains above the background even at 13.6 T.